\pdfoutput=1

\documentclass[conference]{IEEEtran}

\IEEEoverridecommandlockouts

\usepackage{cite}

\ifCLASSINFOpdf
\usepackage[pdftex]{graphicx}
\else
\usepackage[dvips]{graphicx}
\fi

\usepackage[ruled,vlined,linesnumbered]{algorithm2e}
\usepackage[]{caption}
\usepackage[font=footnotesize]{subfig}
\usepackage{stfloats}
\usepackage{amssymb}

\makeatletter
\DeclareRobustCommand*\textsubscript[1]{
	\@textsubscript{\selectfont#1}}
\def\@textsubscript#1{
	{\m@th\ensuremath{_{\mbox{\fontsize\sf@size\z@#1}}}}}
\makeatother

\newcommand{\OM}{HINT}  

\begin{document}
\bstctlcite{IEEEexample:BSTcontrol}

\title{Towards a Realistic Model for Failure Propagation in Interdependent Networks}

\author{\IEEEauthorblockN{Agostino Sturaro\IEEEauthorrefmark{1},
		Simone Silvestri\IEEEauthorrefmark{2},
		Mauro Conti\IEEEauthorrefmark{1}, 
		Sajal K. Das\IEEEauthorrefmark{2}}
	\IEEEauthorblockA{\IEEEauthorrefmark{1} Department of Mathematics, University of Padua,
		email: agostino.sturaro@studenti.unipd.it, conti@math.unipd.it }
	\IEEEauthorblockA{\IEEEauthorrefmark{2}Department of Computer Science, Missouri University of Science and Technology,
		email: \{silvestris, sdas\}@mst.edu}
	
	\thanks{This work is partially supported by the Defense Threat Reduction Agency grant HDTRA1-10-1-0085 and by the NSF grant CNS-1545037.
		Mauro Conti is supported by a Marie Curie Fellowship funded by the European Commission under the agreement No. PCIG11-GA-2012-321980. This work is also partially supported by the TENACE PRIN Project 20103P34XC funded by the Italian MIUR, and by the Project ``Tackling Mobile Malware with Innovative Machine Learning Techniques" funded by the University of Padua.}
}

\maketitle

\begin{abstract}
	Modern networks are becoming increasingly interdependent. As a prominent example, the smart grid is an electrical grid controlled through a communications network, which in turn is powered by the electrical grid. Such interdependencies create new vulnerabilities and make these networks more susceptible to failures. In particular, failures can easily spread across these networks due to their interdependencies, possibly causing cascade effects with a devastating impact on their functionalities.
	
	In this paper we focus on the interdependence between the power grid and the communications network, and propose a novel realistic model, \OM{} (Heterogeneous Interdependent NeTworks), to study the evolution of cascading failures. Our model takes into account the heterogeneity of such networks as well as their complex interdependencies. We compare \OM{} with previously proposed models both on synthetic and real network topologies. Experimental results show that existing models oversimplify the failure evolution and network functionality requirements, resulting in severe underestimations of the cascading failures.
\end{abstract}

\IEEEpeerreviewmaketitle

\section{Introduction}
Modern networks, such as the smart grid and communications networks, are becoming increasingly interdependent \cite{gao_networks_2012, dong_percolation_2012, dong_robustness_2013, huang_robustness_2011, barthelemy_spatial_2010, rosato_modelling_2008, hu_percolation_2011, bashan_percolation_2011}. As an example, the smart grid is an electrical grid monitored and controlled by one or more control centers \cite{locke2010nist} through a communications network. These control centers remotely interact with smart devices, such as: smart meters, Automatic Generation Controls (AGCs) and Phasor Measurement Units (PMUs), which directly operate on the elements of the grid. For instance, AGCs control power plants for energy generation, while PMUs monitor transmission and distribution substations. In turn, such smart devices, and the communications network itself, are powered by the electrical grid, thus generating complex potential feedback loops.

These complex interdependencies may severely affect network functionalities. In particular, a few initial failures may easily spread across the networks due to such interdependencies, possibly causing cascade effects with a devastating impact. A well-known example of such effects occurred in Italy in 2003 \cite{buldyrev_catastrophic_2010, rosato_modelling_2008}. An initial failure in the power network caused the failure of several nodes in the communications network, which generated a cascading effect that left half of the country with no electricity for several hours. A similar event affected the Northeastern and Midwestern United States, and the Canadian province of Ontario, in 2003 \cite{reportUSblackout2003}. These events are not uncommon, as between 2003 and 2012, more than 600 regional outages occurred in the USA, which affected millions of people \cite{economic2013executive}. While little information is available about the nature of these events (i.e., being accidental or maliciously originated), we believe that interdependencies between the smart grid and communications networks should be particularly protected from malicious exploitation.

For the above reasons it is of fundamental importance to study and understand the evolution of failures in interdependent networks. Previous works mainly considered high level abstractions that allow to formulate elegant mathematical frameworks, generally based on percolation theory\cite{gao_single_2014}. Unfortunately, these approaches overlook several features of these systems, and assume homogeneity of network elements, unrealistic failure propagation conditions and even unrealistic topological constraints \cite{gao_networks_2012, dong_percolation_2012, huang_robustness_2011, barthelemy_spatial_2010, hu_percolation_2011, bashan_percolation_2011}. As a result, these approaches, although mathematically elegant, cannot accurately predict the dynamics of failure evolution in real networks.

In this paper we propose a realistic model, \OM{} (Heterogeneous Interdependent NeTwork), to study the evolution of cascading failures in interdependent networks. In particular, we focus on the power grid, the communications network, and their interdependencies. Unlike previous works, we address the heterogeneity of the structure of the power grid by considering its nodes to be functionally separated in three categories: \textit{generation}, \textit{transmission} and \textit{distribution}. Similarly, we differentiate nodes in the communications network, identifying the \textit{control centers} that remotely operate the power grid, and the \textit{relay} nodes which are only responsible for data communication. We define a set of {\em logical} and {\em physical} interdependencies as well as a set of conditions to determine the evolution of failures across networks.

We further compare our model, \OM{}, with two previously proposed approaches, using realistic synthetic topologies, as well as real topologies from the Minnesota Power grid \cite{minnesota_power} and communications network \cite{auroraNet}. Our experimental results show that existing approaches largely underestimate the effects of cascading failures, due to their simplifying assumptions. On the contrary, our model is able to capture the complex functionalities and interactions of these systems and represents an important step towards an accurate model of cascading failures in interdependent networks.

In summary, the major contributions of this paper are:
\begin{itemize}
	\item We propose \OM{}, a realistic model for failure propagation in interdependent power and communications networks, which takes into account their heterogeneity, as well as their logical and physical interdependencies.
	\item We define a set of conditions governing the evolution of failures that account for network heterogeneity and complex interdependencies.
	\item We compare \OM{} with previous models on synthetic and real networks. Results show that our model better predicts the evolution of cascading failures.
\end{itemize}

\section{Existing models for failure propagation}
Several models have been proposed to study the spreading of failures in interdependent networks \cite{buldyrev_catastrophic_2010, buldyrev_interdependent_2011,huang_balancing_2013,huang_small_2015}. Most of these works focus on the power grid and the communications network, and share a common framework according to which both networks can be represented as undirected graphs. In particular, the power grid is modeled as a graph $G_{pow} = (V_{pow}, E_{pow})$, while the communications network as a graph $G_{com} = (V_{com}, E_{com})$. Dependencies between nodes in the different networks are represented as a set $E_{dep}$ of edges between nodes in the two graphs. 
Due to the dependencies defined by the set $E_{dep}$, a node failure in one network may cause the failure of other nodes within the same network, as well as on the other network. Existing models for failure propagation differ in how they assign interdependencies between networks and in how they define the conditions that a node must meet in order to be considered operational.

In the following, we describe two existing models that we use for comparison with the realistic model proposed in this paper.
The first model has been one of the first proposed in this context, in the pioneering work of Buldyrev et al. \cite{buldyrev_catastrophic_2010}. We refer to this model as the {\em Uniform model}. It is a relatively simple model, based on several simplifying assumptions. In contrast, the second model has been recently proposed \cite{huang_small_2015}, and addresses more general scenarios and more complex failure propagation conditions. We refer to this model as the {\em Small Clusters} model. Figure \ref{fig:Legend} summarizes the notations adopted for nodes and links.

\subsection{Uniform model}\label{sec:uniform_model}
The Uniform model \cite{buldyrev_catastrophic_2010} assumes that all nodes, in both the power and communications networks, are homogeneous in terms of their roles in the network and in the failure propagation process. Because of this uniform behavior, we refer to this model as the {\em Uniform model}.

\noindent
{\bf Interdependencies:} This model assumes the so called \textit{one-to-one} dependency, that is, a node $u$ in a network depends on exactly one node $v$ in the other network, and, vice-versa, $v$ depends only on $u$. To represent this, we consider the edges in $E_{dep}$ to be undirected, and the inter-degree of each node to be exactly one. Figure \ref{subfig:ex_1_uniform_start} shows the structure of the Uniform model. The roles of the represented nodes have no influence on this model.

Both assumptions mentioned above considerably simplify the structure of real networks and their interdependencies. In fact, real networks are highly heterogeneous, and their interdependencies can be multiple, directed and asymmetrical.

\noindent
{\bf Failure propagation:} According to the Uniform model, a node $u$ is considered \textit{operational} if it belongs to the {\em giant component} of its network and if the node $v$ on which it depends, i.e., $(u,v) \in E_{dep}$ is also operational. 

Let $GC(G_{pow})$ and $GC(G_{com})$ be the set of currently operational nodes in the giant component of the power and communications network, respectively. Let us consider a node $u \in V_{pow}$ in the power network. This node fails if (i) $u \notin GC(G_{pow})$ (intra-network failure), and (ii) $\nexists (u, v) \in E_{dep}$ s.t. $v \in GC(G_{com})$ (inter-network failure). Similar conditions can be easily defined for nodes in the communications network.

Given the above failure propagation conditions, we provide in Algorithm Simulation the scheme used to determine the failure propagation. We denote by $F^{0}$ the initial set of failed nodes at time $t=0$, and we assume for simplicity that nodes initially fail in the power network\footnote{A similar approach can be used to consider nodes initially failed in the communications network, or in both networks.}, i.e., we assume $F^{0} \subseteq V_{pow}$. Additionally, we denote by $F_{pow}^{t}$ and $V_{pow}^{t}$ the set of nodes that at round $t$ are failed and operational, respectively. We use a similar notation, $F_{com}^{t}$ and $V_{com}^{t}$, for the communications network.

Given the initial failed nodes, the failure conditions are checked in phases. In the first phase (lines 6-7), the algorithm checks the new intra-network failures in the power network, i.e., if there are nodes that no more belong to the giant component. In the second phase (lines 8-9), it checks if there are some additional inter-network failures, i.e., if some power nodes depend on failed communications nodes, and hence they also fail.
The last two phases (lines 10-13) perform similar checks for nodes on the communications network. The algorithm iterates until the cascading failure effects converge and no more failures occur.

\begin{algorithm}[h]
	\caption{Simulation}
	\label{alg:simulation}
	\SetAlgoRefName{Simulations}

	\small
	\KwIn{$G_{pow}(V_{pow}, E_{pow})$ power grid\\
		$G_{com}(V_{com}, E_{com})$ communications network\\
		$F^{0} \subseteq V_{pow}$ initial set of failures}
	\BlankLine
	\KwOut{Graphs at steady state, after cascading failures}
	\BlankLine

	\tcp{initialization}
	$t \leftarrow 0$\;
	$V_{pow}^{t} \leftarrow V_{pow}^{t-1} \setminus F^{0}$\;
	$V_{com}^{t} \leftarrow V_{com}$\;
	\BlankLine

	\While{additional failures are possible}{

		$t \leftarrow t+1$\;
		\BlankLine

		\tcp{Intra-network failures for $G_{pow}$}
		$F_{pow}^{t}$ = nodes in $V_{pow}^{t-1}$ not in $GC^{t}(G_{pow})$\;
		$V_{pow}^{t} \leftarrow V_{pow}^{t-1} \setminus F_{pow}^{t}$\;
		\BlankLine

		\tcp{Inter-network failures for $G_{pow}$}
		$F_{pow}^{t}$ = nodes in $V_{pow}^{t}$ with no support in $G_{com}$\;
		$V_{pow}^{t} \leftarrow V_{pow}^{t} \setminus F_{pow}^{t}$\;
		\BlankLine

		\tcp{Intra-network failures for $G_{com}$}
		$F_{com}^{t}$ = nodes in $V_{com}^{t-1}$ not in to $GC^{t}(G_{com})$\;
		$V_{com}^{t} \leftarrow V_{com}^{t-1} \setminus F_{com}^{t}$\;
		\BlankLine

		\tcp{Inter-network failures for $G_{com}$}
		$F_{com}^{t}$ = nodes in $V_{com}^{t}$ with no support in $G_{pow}$\;
		$V_{com}^{t} \leftarrow V_{com}^{t} \setminus F_{com}^{t}$\;

	}
\end{algorithm}

\subsection{Small clusters model}
The Small Clusters (SC) model \cite{huang_small_2015} is more advanced and takes into account different roles of nodes in the communications network, more complex network interdependencies and failure propagation schemes. In particular, this model introduces two types of nodes in the communications network, \textit{control nodes} and \textit{relay nodes}. Control nodes are responsible for controlling the functionality of the nodes in the power grid, while relay nodes provide the communications infrastructure of the communications network. We refer to the set of control nodes as $V_{cc}$, and to the set of relay nodes as $V_{rl}$. Therefore, $V_{com} = V_{cc} \cup V_{rl}$. Figure \ref{subfig:ex_1_sc_start} shows the structure of the SC model. Only the roles of nodes in the communications network are considered.

\noindent
{\bf Interdependencies:} The model adopts an interdependency structure known as the \textit{k-n} dependency, initially proposed in \cite{huang_balancing_2013}. According to this model each node in $G_{pow}$ depends on \textit{k} control nodes in $G_{com}$, while every control node in $G_{com}$ supports \textit{n} nodes in $G_{pow}$. Additionally, every node in $G_{com}$ depends on \textit{one} power node in $G_{pow}$ for power supply. Dependencies are directional and asymmetric, so if a node $u$ depends on a node $v$, it does not imply that $v$ depends on $u$.

Since dependencies are directional (arcs instead of edges), two different sets of dependency arcs are introduced, $A_{ctrl} \subseteq V_{pow} \times V_{com} $ and $A_{energy} \subseteq V_{com} \times V_{pow}$, where $\times$ is the Cartesian product operation. An arc $(u,v) \in A_{ctrl}$ means that the power node $u$ is controlled by the control center $v$. Similarly, an arc $(u,v) \in A_{energy}$ means that a communications node $u$ is powered by a node $v$ of the power network.

\noindent
{\bf Failure propagation:} The SC model is based on the observation that the giant component, as a criteria to determine if a node is functional, is not representative of the operation of real systems. The authors point out in \cite{huang_small_2015} that sufficiently large components can also be operational, although disconnected from the giant component. For this reason, the model introduces a threshold $\Delta$ on the size of a component, above which the nodes are considered to be able to operate. In addition, such nodes also need to have at least an interlink to the other network. As a result, some autonomous clusters can keep operating although they become disconnected from the giant component.

In the SC model, the propagation of failures across networks is determined using a scheme similar to the Uniform model. We do not provide the pseudo-code due to space limitations. However, it can be easily obtained from Algorithm Simulation by substituting the failure conditions of the Uniform model with those of the SC model.

The SC model significantly improves previous approaches, however it still overlooks several important aspects of real systems. In particular, all nodes in the power networks are considered homogeneous, while in real power system we observe nodes which produce energy, transfer energy, and distribute energy. In addition, the size of a component alone is not sufficient to determine its capability to operate. For instance, a group of substations will not be able to distribute any energy if it is not connected to any power plant. Finally, the SC model only considers logical dependencies of power nodes with the control center. It does not consider that such control can only be realized if the power nodes can access the communications network through a relay node to communicate with their control center. In the next section we introduce our realistic failure propagation model, which takes into account these aspects of real systems.

\section{A realistic model for failure propagation}
In this section we introduce \OM{}, our proposed model for realistic failure propagation in interdependent power and communications networks. Unlike the Uniform and SC models, \OM{} specifically addresses the heterogeneity of the structure of both networks, and differentiates between {\em logical} and {\em physical} interdependencies. Based on these features, we define a set of conditions to determine the spreading of intra- and inter-network failures.

\noindent
{\bf Network heterogeneity:} We consider nodes in the power grid $G_{pow}$ to be functionally separated in three categories\footnote{We do not address the customer level of the power grid in this work, since it has marginal effects on the failure propagation. Nevertheless, the model can be easily extended to also include such level.}: \textit{generation} nodes $V_{gen}$, \textit{transmission} nodes $V_{ts}$, and \textit{distribution} nodes $V_{ds}$. Generation nodes, such as power plants, produce electricity. Transmission nodes connect power lines and are used for switching current. Finally, distribution nodes are substations containing transformers that step down the voltage and distribute electricity to the final users.

Similarly, we differentiate nodes in the communications network $G_{com}$ considering control centers $V_{cc}$, that remotely operate the power grid, and relay nodes $V_{rl}$ which are only responsible for data communication. Figure \ref{subfig:ex_1_realistic_start} shows the structure of \OM{}, the notation is explained in Figure \ref{fig:Legend}. The roles of all the represented nodes are taken into account by our model.

\noindent
{\bf Interdependencies:} \OM{} specifically considers {\em logical} and {\em physical} interdependencies between the power and the communications network. Physical dependencies occur between the nodes in the communications network and nodes in the power network. In particular, a node in the communications network has a physical dependency with the nodes in the power network from which it receives power. Differently from previous works, we impose that only distribution nodes can provide power to the communications nodes, as is the case in real systems. We represent such dependencies with the set of directional inter-arcs $A_{energy}$.

Logical dependencies occur between nodes in the power network and control centers in the communications network. These dependencies represent the control that power nodes need to operate, and are represented by the set of arcs $A_{ctrl}$.

Finally, we introduce a new kind of physical dependencies between power nodes and communications nodes. In fact, although a power node is controlled by a control center, this control is only possible if the power node can access the communications network. For this reason, in our model each node in the power network has a physical dependency with a relay node, which is used to access the communications network. We represent such dependencies with the set of directional inter-arcs $A_{info}$. Note that the dependencies in $A_{ctrl}$ indicate the required logical information-exchange with the control center, while the dependencies in $A_{info}$ refer to actual links used to exchange such information.

\noindent
{\bf Failure propagation:} The failure propagation scheme in \OM{} takes into account the heterogeneity of nodes as well as the physical and logical dependencies described above. In the following, we denote by $op(u)$ the fact that node $u$ is currently operational, and by $\neg op(u)$ otherwise. Additionally, given two nodes, $u$ and $v$, in the same network, we introduce the function $path(u, v)$. This function returns a set of nodes that are currently operational and constitute a path from $u$ to $v$ in their network, if it exists, and $\emptyset$ otherwise.

We consider a node $u$ in the communications network to \textit{fail} only if none of the distribution substations it draws power from is operative. That is, $u$ fails if $\forall v$ s.t. $(u, v) \in A_{info}$, $\neg op(u)$. Note that, as in real systems, in our model the functionality of a node in the communications network does not depend on the size of the component it belongs to, as long as it receives power.

According to \OM{}, a node $v$ in the power grid may fail for several reasons. Node $v$ fails if all the control centers that are responsible for its operation have failed, that is $\forall u$ s.t. $(v, u) \in A_{ctrl}$, $\neg op(u)$. Additionally, although some control centers may be operational after the failure, such control cannot be performed if $v$ cannot access the communications network because the relay nodes it uses for this purpose have failed. For this reason we also introduce the following condition, $v$ fails if $\forall u$ s.t. $(v, u) \in A_{info}$, $\neg op(u)$.

The access to the communications network is necessary but not sufficient to ensure the operation of $v$. In fact, for a control center to be able to properly control $v$, there must be connectivity between the control center and at least a relay node that $v$ uses to access the communications network. Formally, $v$ fails if $\forall u,q$ s.t. $(v, u) \in A_{info}$ and $(v, q) \in A_{ctrl}$, $path(u, q) = \emptyset$.

Finally, a node $v$ in the transmission and distribution network can be operational only if it receives power from a generator. We model this by allowing $v$ to be functional only if there exists at least a path in the power network that connects $v$ to a generator which is currently operational. In other words, a distribution or transmission node $v$ fails if $\forall q \in V_{gen}$ s.t. $op(q)$, $path(v, q) = \emptyset$.

\section{An example of failure propagation}\label{sec:examples}
\begin{figure}[!h]
	\hspace*{\fill}
	\includegraphics[width=0.9\columnwidth]{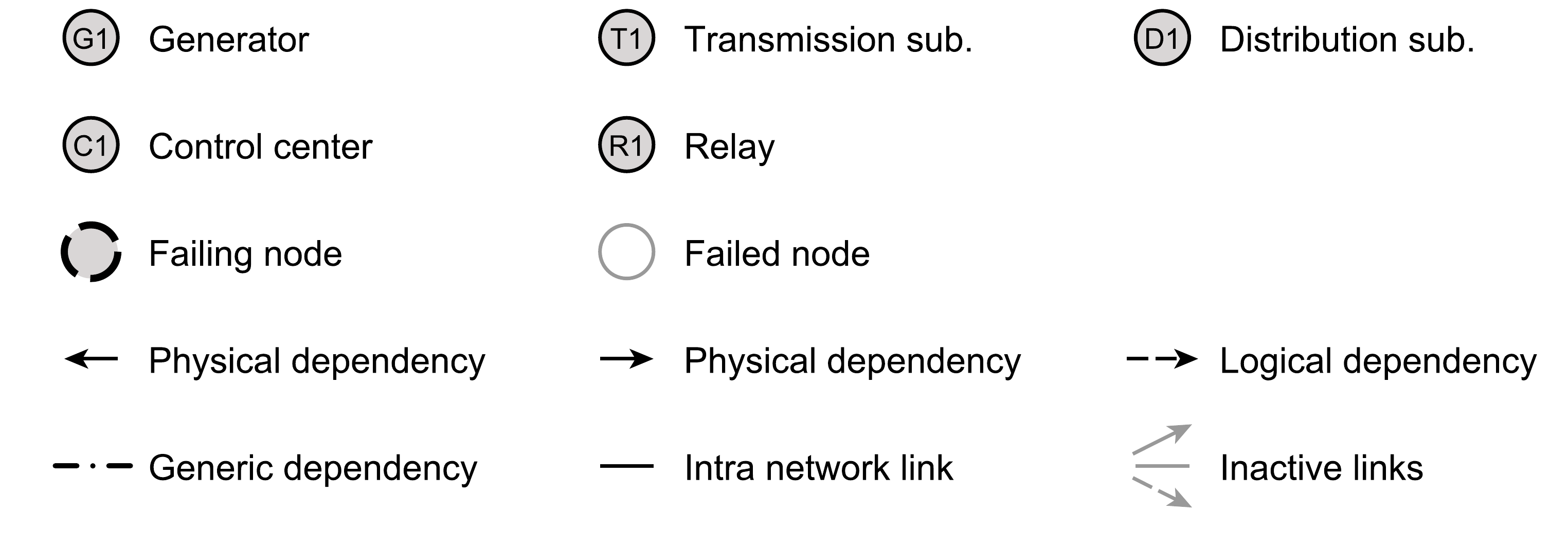}
	\hspace*{\fill}
	\caption{Graphical notation used for nodes and links in Figures.}
	\label{fig:Legend}
\end{figure}

\begin{figure}[!h]
	\vspace{-1em}
	\hspace*{\fill}
	\subfloat[\label{subfig:ex_1_realistic_start}]{%
		\includegraphics[width=0.48\columnwidth]{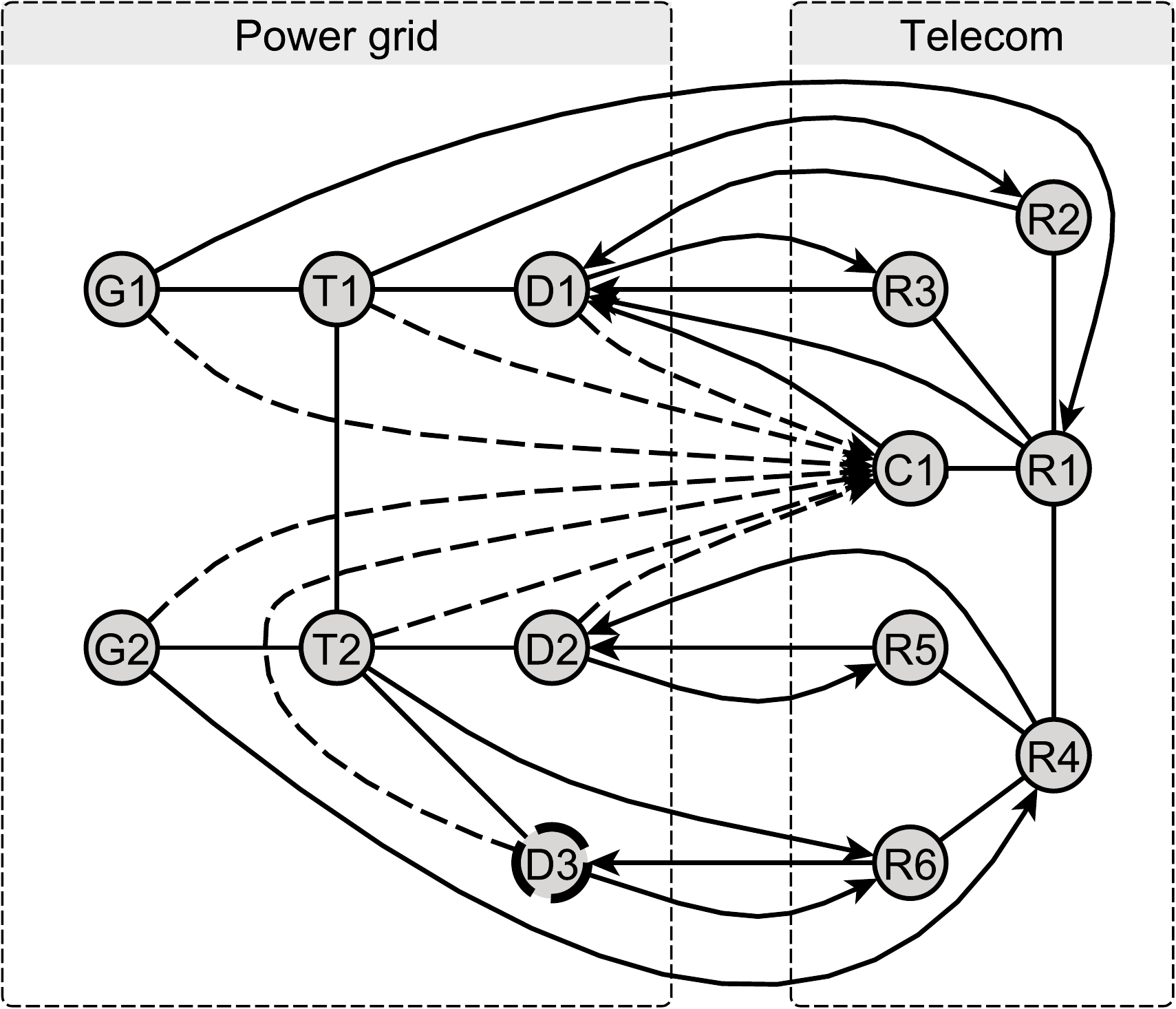}
	}\hfill
	\subfloat[\label{subfig:ex_1_realistic_stop}]{%
		\includegraphics[width=0.48\columnwidth]{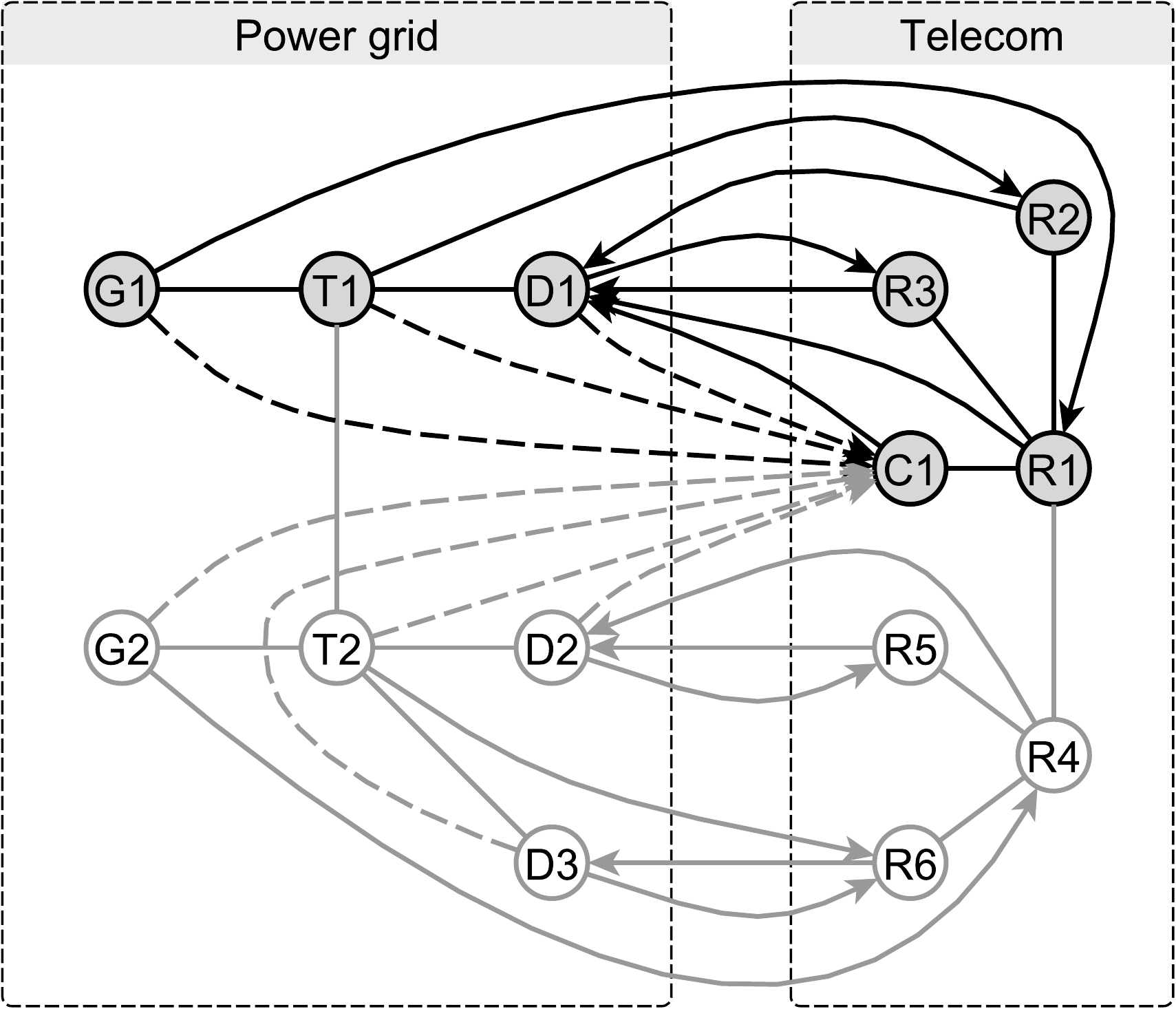}
	}\hspace*{\fill}
	\caption{Scenario 1, proposed realistic model \OM{}: (a) initial configuration, (b) final stable state.}
	\label{fig:example_1_our}
\end{figure}

In this section we present an example to compare the evolution of cascading failures under the Uniform, SC and \OM{} models.
Figure \ref{fig:Legend} summarizes the notations adopted for nodes and links. Figure \ref{subfig:ex_1_realistic_start} shows the considered example.

The scenario includes a power grid with two generators, two transmission substations, and three distribution nodes. The communications network, on the other hand, has a single control center and several relay nodes that enable communications with the nodes in the power grid. All the nodes in the power grid have a logical dependency with the control center, but they also have physical dependencies with relay nodes in the communications network to realize such control. Finally, nodes in the communications network also have physical dependencies on the nodes in the distribution level of the power grid, from which they receive power supply.

In this example, we consider an attack that disables the distribution substation \textit{D3}. According to our model, \OM{}, the apparently innocuous failure of \textit{D3} has devastating effects. In particular, the failure of \textit{D3} causes the failure of the relay \textit{R6}, since \textit{D3} provides energy to it. However, \textit{R6} is used by \textit{T2} to access the communications network, and hence \textit{T2} also fails. This disconnects the generator \textit{G2}, leading to several other cascading effects on both networks and terminating in the configuration shown in Figure \ref{subfig:ex_1_realistic_stop}.

\begin{figure}[!h]
	\vspace{-1em}
	\hspace*{\fill}
	\subfloat[\label{subfig:ex_1_sc_start}]{%
		\includegraphics[width=0.48\columnwidth]{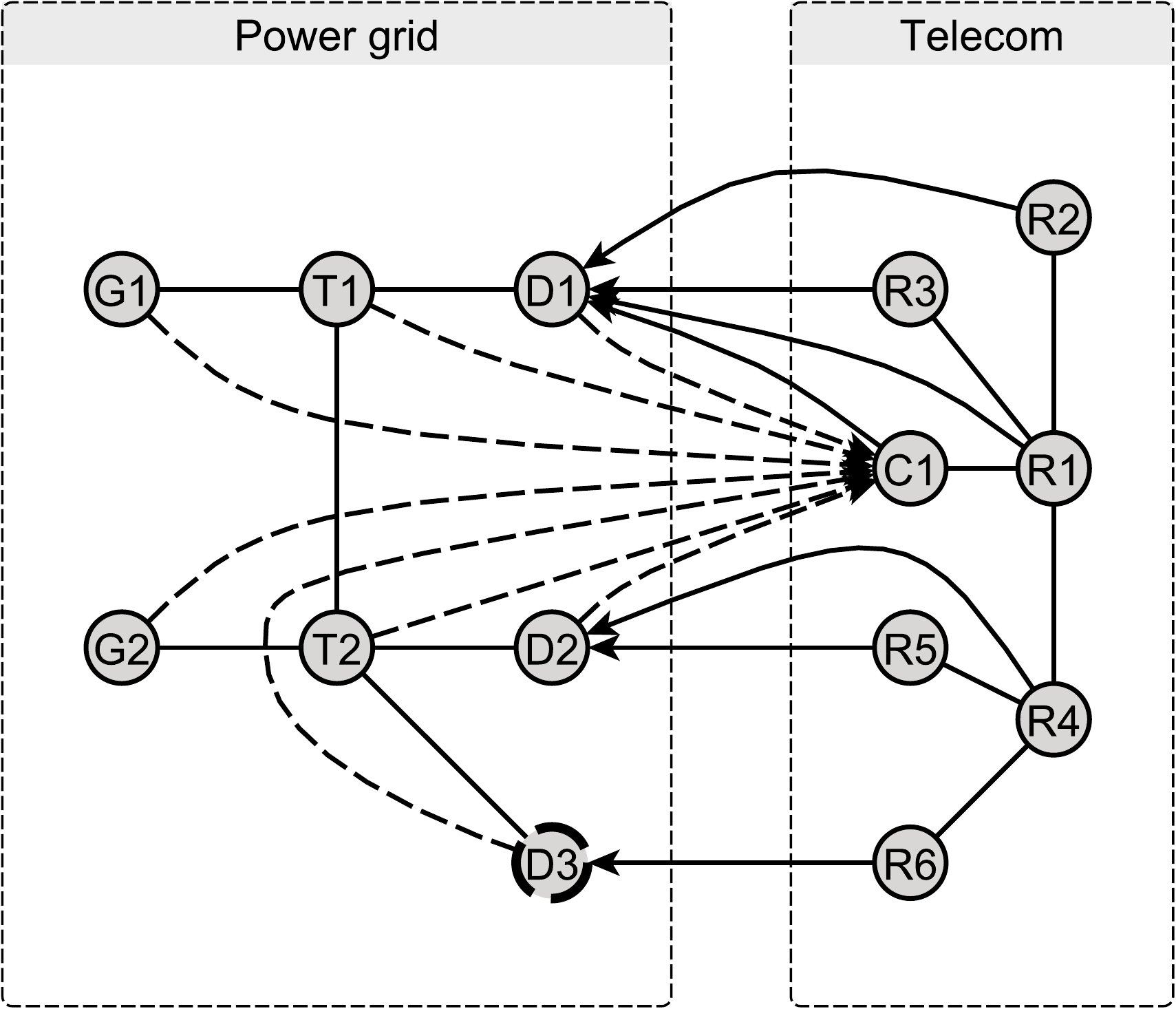}
	}\hfill
	\subfloat[\label{subfig:ex_1_sc_th_4_stop}]{%
		\includegraphics[width=0.48\columnwidth]{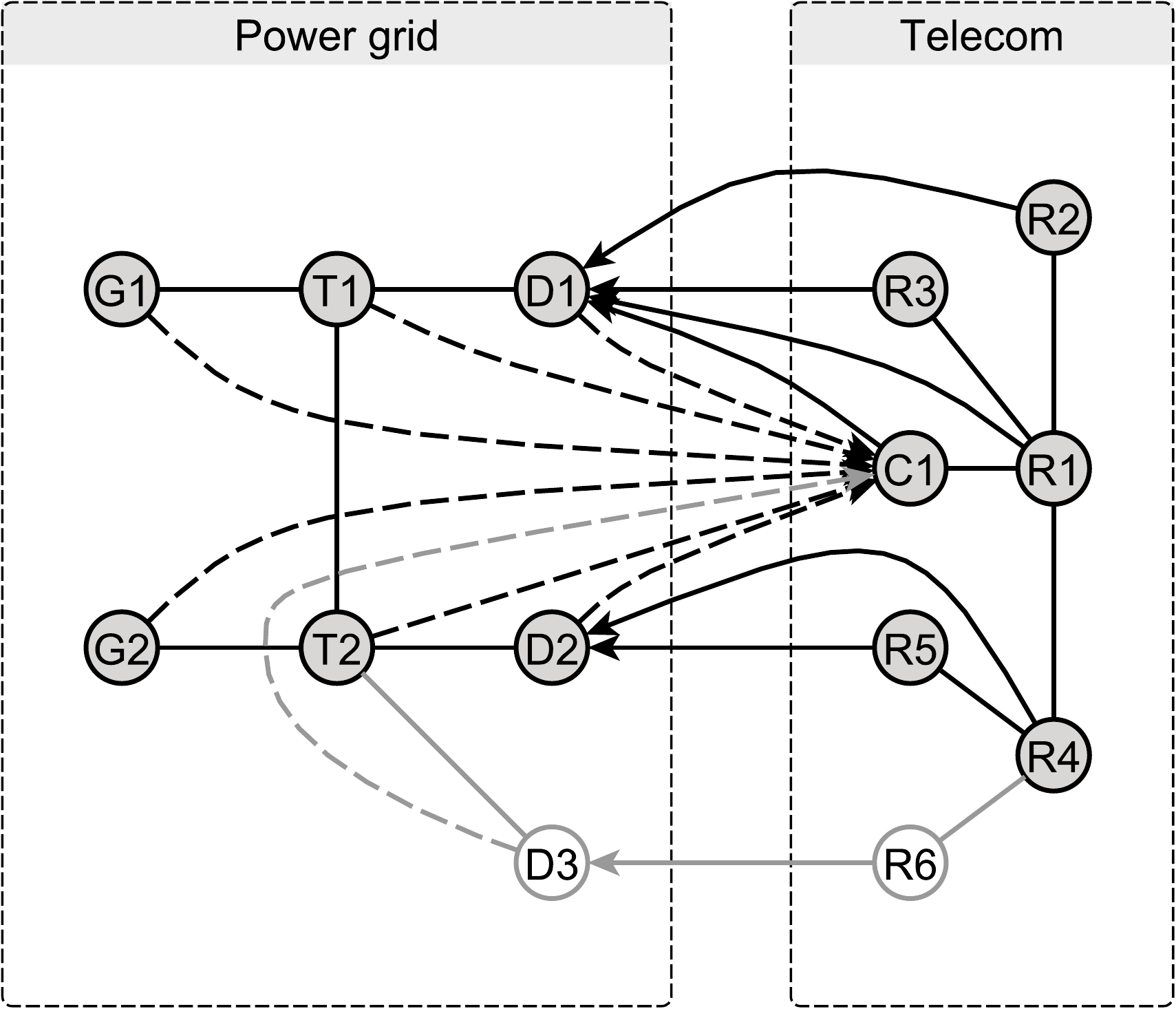}
	}\hspace*{\fill}
	\caption{Scenario 1, SC model: (a) initial configuration, (b) final stable state $\Delta = 4$.}
	\label{fig:example_1_sc}
\end{figure}

We consider the same scenario under the SC model in Figure \ref{fig:example_1_sc}. We set $\Delta = 4$. Figure \ref{subfig:ex_1_sc_start} shows the initial configuration. It is important to note that the SC model {\em does not} consider the physical dependencies between power nodes and the relay nodes used to access the communications network. As a result, this model does not fully capture the cascade generated by the failure of $D3$. In particular, the failure of \textit{D3} only causes the failure of \textit{R6}. No additional failure occurs, since the dependency of \textit{T2} from \textit{R6} is not considered by the model, and the surviving components are larger than $\Delta$. As a result, the model underestimates the final failures, as shown in Figure \ref{subfig:ex_1_sc_th_4_stop}. In this example, any value of $\Delta$ less than 6 leads to the same final result.

\begin{figure}[!h]
	\vspace{-1em}
	\hspace*{\fill}
	\subfloat[\label{subfig:ex_1_uniform_start}]{%
		\includegraphics[width=0.48\columnwidth]{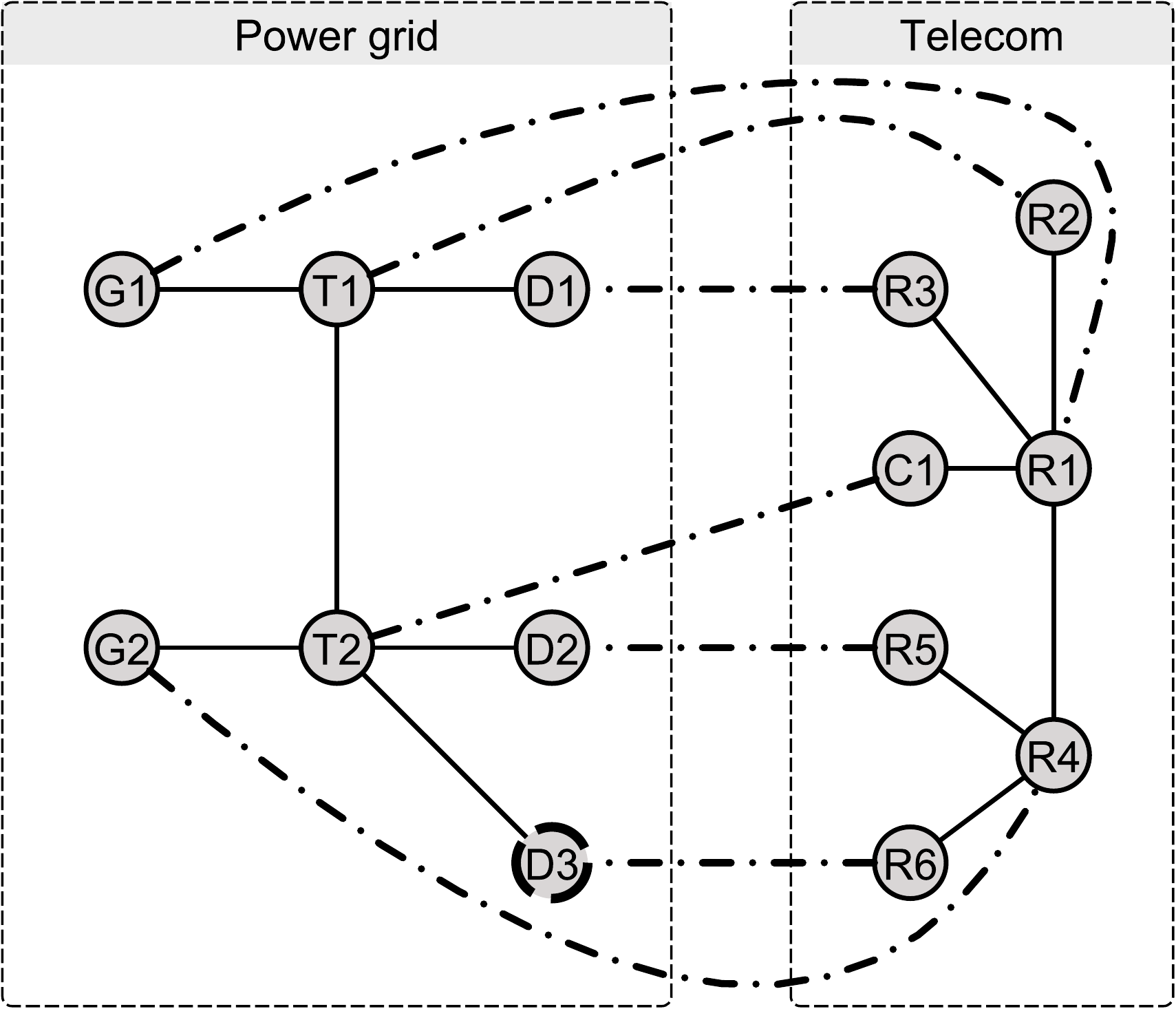}
	}\hfill
	\subfloat[\label{subfig:ex_1_uniform_stop}]{%
		\includegraphics[width=0.48\columnwidth]{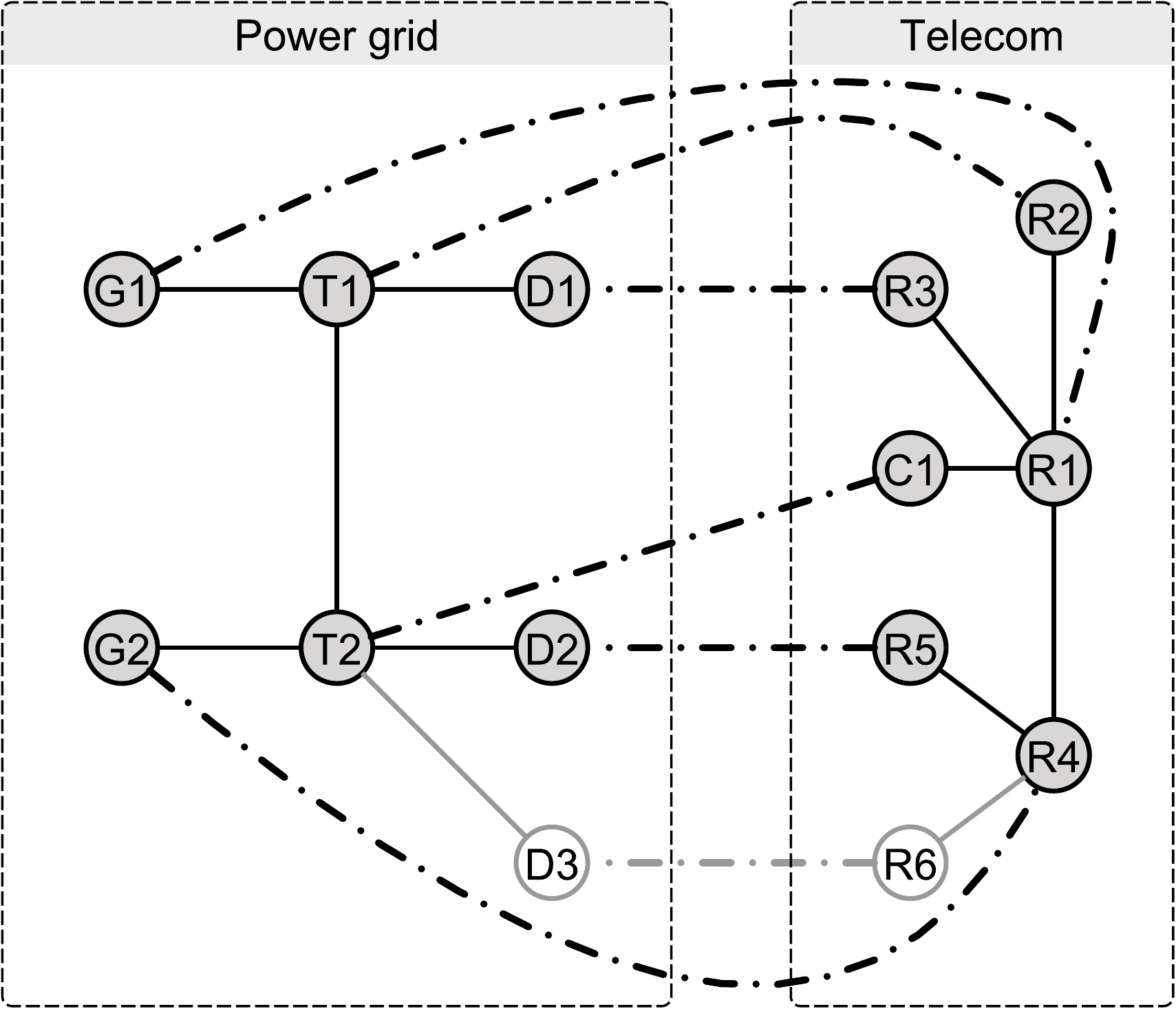}
	}\hspace*{\fill}
	\caption{Scenario 1, Uniform model: (a) initial configuration, (b) final stable state.}
	\label{fig:example_1_uniform}
\end{figure}

Figure \ref{fig:example_1_uniform} shows the same scenario under the Uniform model. We recall that this model imposes one-to-one dependencies. As a result, we cannot include all the actual dependencies. In order to keep as many as possible, we use a maximum matching algorithm. First we create a bipartite graph consisting of power grid nodes, communications network nodes and their interdependencies. Then we run the algorithm to select the largest set of edges such that no two edges share a common vertex.
This way, we can keep the maximum number of dependencies without violating the one-to-one requirement of the model and obtain the initial configuration shown in Figure \ref{subfig:ex_1_uniform_start}. Several dependencies are necessarily left out, which reduces the accuracy of the model. In particular, the failure of $D3$ only causes the failure of $R6$, because after these failures all nodes are still part of the giant component of their respective networks. Therefore, the Uniform model also underestimates the final failures, as shown in Figure \ref{subfig:ex_1_uniform_stop}.

\vspace{-1em}
\section{Experimental results}
In this section we compare our model, \OM{} with the Uniform and SC models through simulation experiments. We use realistic synthetic network topologies as well as real network topologies.
The simulator we developed to run these experiments uses the NetworkX library \cite{networkx_paper}, we called it TiedNets and made it freely available on GitHub \cite{our_simulator}.

The synthetic power network is created according to the realistic RT-nested-Smallworld model proposed in \cite{wang_generating_2010}.
This model generates a given number of well connected {\em subnetworks}, which are subsequently interconnected. We generated power networks with 1000 nodes and 20 subnetworks. The average node degree is 4, which is a typical value for real networks such as the United States Northeastern power network and the European power network \cite{wang_generating_2010}. Since the model does not specify how to classify nodes in the resulting power grid, we used 100 generators, 270 transmission substations and 630 distribution substations. We assigned such roles so that each subnet has a similar proportion of nodes in each category.

We generated a synthetic communications network using the Barab\'asi-Albert model \cite{barabasi_emergence_1999}. This model is well known to generate scale-free networks with power-law degree distribution, and topological structures similar to real communications networks such as the Internet. In this model, nodes are added iteratively. Each newly added node connects to $m$ existing nodes. The probability of connecting to a node is proportional to its degree. In our experiments we set $m = 3$. We consider a special node in the network as a control center, while the remaining nodes are relay nodes. Also in this case we generated a network of 1000 nodes.

We interconnect the two synthetic networks as follows. Each node in the power network has a logical dependency from the control center and also a physical dependency from a relay node in the communications network. Each node in the communications network has a dependency with a node in the power network from which it receives energy. All physical dependencies are randomly assigned.

In order to validate \OM{} and compare it with the Uniform and SC models, we also use real network topologies. In particular, we use the Minnesota Power grid \cite{minnesota_power}. Our data provides extremely detailed topological information, with all power lines down to $69$ kV and most substations of the State. We want to highlight that this is the first work that makes use of such detailed data. 
The network has 1022 nodes. In addition, we include generator nodes using the list of power plants provided in \cite{mn_power_plants} and their positions \cite{enipedia}. Overall, we identify 69 power plants that include all power plants in the state with a production higher than or equal to $14$ MW. The total number of nodes in the network is thus 1091.

The real communications network is the fiber optics network of Minnesota provided by AuroraNet \cite{auroraNet}. The network has 681 nodes, to which we attach a node as a control center, located in the vicinity of the most populated areas. 

We interconnect the real networks as follows. All nodes in the power network have a logical dependency with the control center. We assign physical dependency using a geographic criterion. In particular, power nodes depend on the nearest relay node, while communications nodes depend on the nearest distribution substation.

In all the following experiments, we average the results of several runs and show the achieved standard deviation.

\subsection{Random attacks}
In the first set of experiments we compare the models under random attacks. In particular, we initially fail a given number of nodes selected uniformly at random, and then compare the final number of nodes that have failed under the considered models after the cascade.

\begin{figure}[]
	\vspace{-1em}
	\hspace*{\fill}
	\subfloat[\label{subfig:random_atk_synthetic}]{%
		\includegraphics[width=0.48\columnwidth]{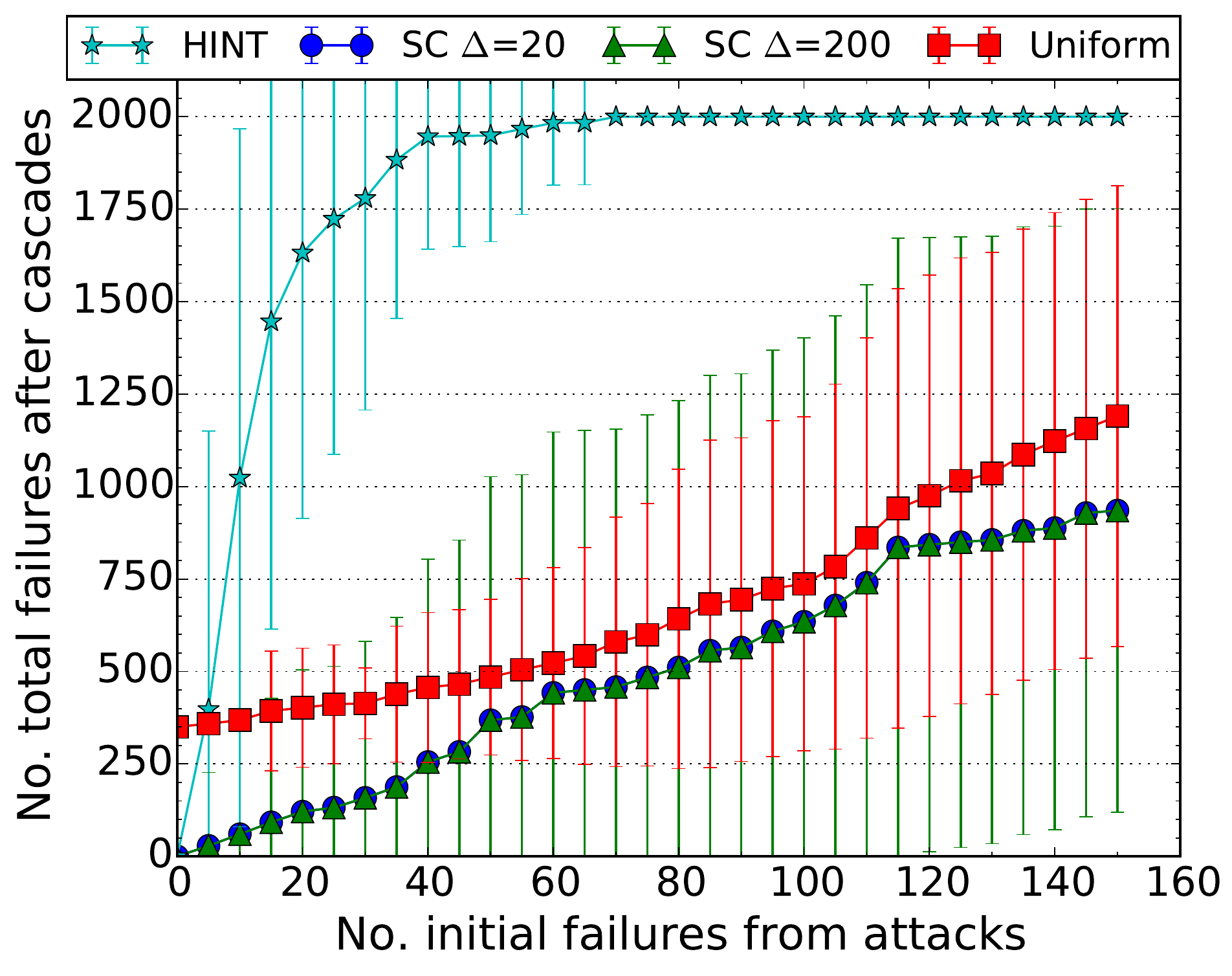}
	}\hfill
	\subfloat[\label{subfig:random_atk_real}]{%
		\includegraphics[width=0.48\columnwidth]{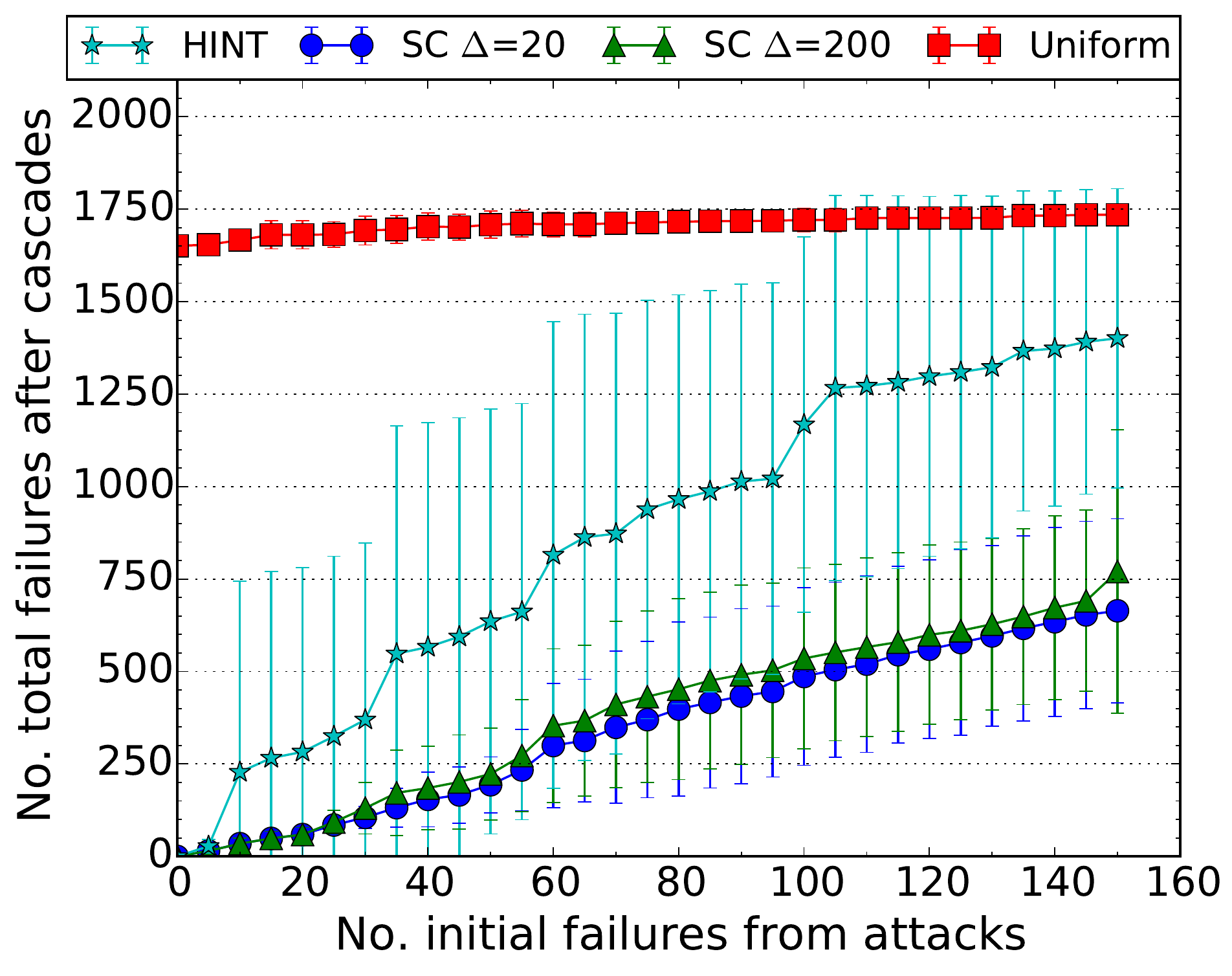}
	}\hspace*{\fill}
	\caption{Random attacks (a) on synthetic networks, and (b) on real networks.}
	\label{fig:random_atk}
\end{figure}

Figure \ref{subfig:random_atk_synthetic} shows the results on synthetic networks. Our proposed model demonstrates the significant vulnerability of the considered networks against random attacks. Attacking as few as $2.5\%$ of the total nodes may cause a complete outage of both the power and the communications networks. The heterogeneity of roles in both network, as well as the complex rules that define the nodes' ability to operate, unveil the extreme vulnerability of these networks when coupled together. These results highlight the importance of considering network interdependencies when designing future smart grids, to improve their robustness and reliability.

On the contrary, the other two models significantly underestimate the size of the cascade. In particular, for the SC model we use two settings for the threshold $\Delta$, such as $\Delta = 20$ and $\Delta = 200$. The first setting is used in \cite{huang_small_2015}, however, that paper does not provide a methodology to set $\Delta$. For comparison, we also considered $\Delta = 200$, that imposes a significantly larger component size for nodes to correctly operate. Nevertheless, as Figure \ref{subfig:random_atk_synthetic} shows, both setting behave similarly. This is due to the high connectivity of both the power and the communications networks, which prevents network partitioning due to random attacks \cite{albert_error_2000}, and thus prevents further spreading of failures. The only case of large scale failures occurs when the control center fails due to the initial attack. This justifies the large variability of the results shown by the error bars.

The Uniform model apparently performs better than the SC model, being able to predict the failure of a higher number of nodes. However, this is due to the instability of the system that results from the one-to-one dependency assignment requirement by the model. Although we maximize the number of dependencies thanks to the use of the maximum matching (discussed in Section \ref{sec:examples}), a perfect matching may not be possible, therefore leaving some nodes with no dependency and unable to operate.

Figure \ref{subfig:random_atk_real} shows the results for the random attacks in real networks. \OM{} shows that real networks are more robust compared to the random networks. This is due to the geographical assignment of the dependencies, which closely resembles how nodes are interconnected in reality. In particular, the geographic assignment tends to localize failures in the network, making it harder for few initial failures to achieve a global cascading effect. On the contrary, in the synthetic networks interdependencies are assigned at random, and a failure originated in one part of a network can easily spread to remote parts of the other network. These results highlight the importance of using real network topologies for studying the evolution of cascading failures, which may reveal unexpected effects not captured by synthetic models.

The SC model in this case also underestimates the final size of the cascade. For these experiments we consider two setting for the threshold, $\Delta =$ 20 and 200. Although the two settings behave slightly differently, in both cases the model underestimates the effects of the failure. Overall, the inability of the SC model to include physical dependencies and heterogeneity of the power network, significantly penalizes its accuracy and causes underestimation of the final size of the cascade.

Using real network topologies, the Uniform model is critically unstable, due to the inability of finding a suitable maximum matching. Because of the cascade of failures caused by the large number of unsupported nodes, most nodes fail without being attacked.

\subsection{Targeted attacks}
In the second set of experiments, we consider how \OM{} reacts to targeted attacks. Several previous works consider similar attacks \cite{dong_percolation_2012, huang_robustness_2011}, generally based on the node degree. However, these works are based on simplified models similar to the Uniform model, preventing a fine granularity of the attack. In our work, we consider targeted attacks which specifically take into account the role of nodes in the network and their interdependency with the other network. In particular our attacks focus on the power grid and test four kinds of attacks. The first attack initially fails distribution nodes in the power grid, in decreasing order of their inter-degrees (the number of nodes that depend on them in the other network).
The other three attacks, instead, consider distribution, transmission and generation nodes, respectively, in decreasing order of their intra-degree (the number of nodes they are connected to in their own network).

\begin{figure}[]
	\hspace*{\fill}
	\includegraphics[height=0.46\columnwidth]{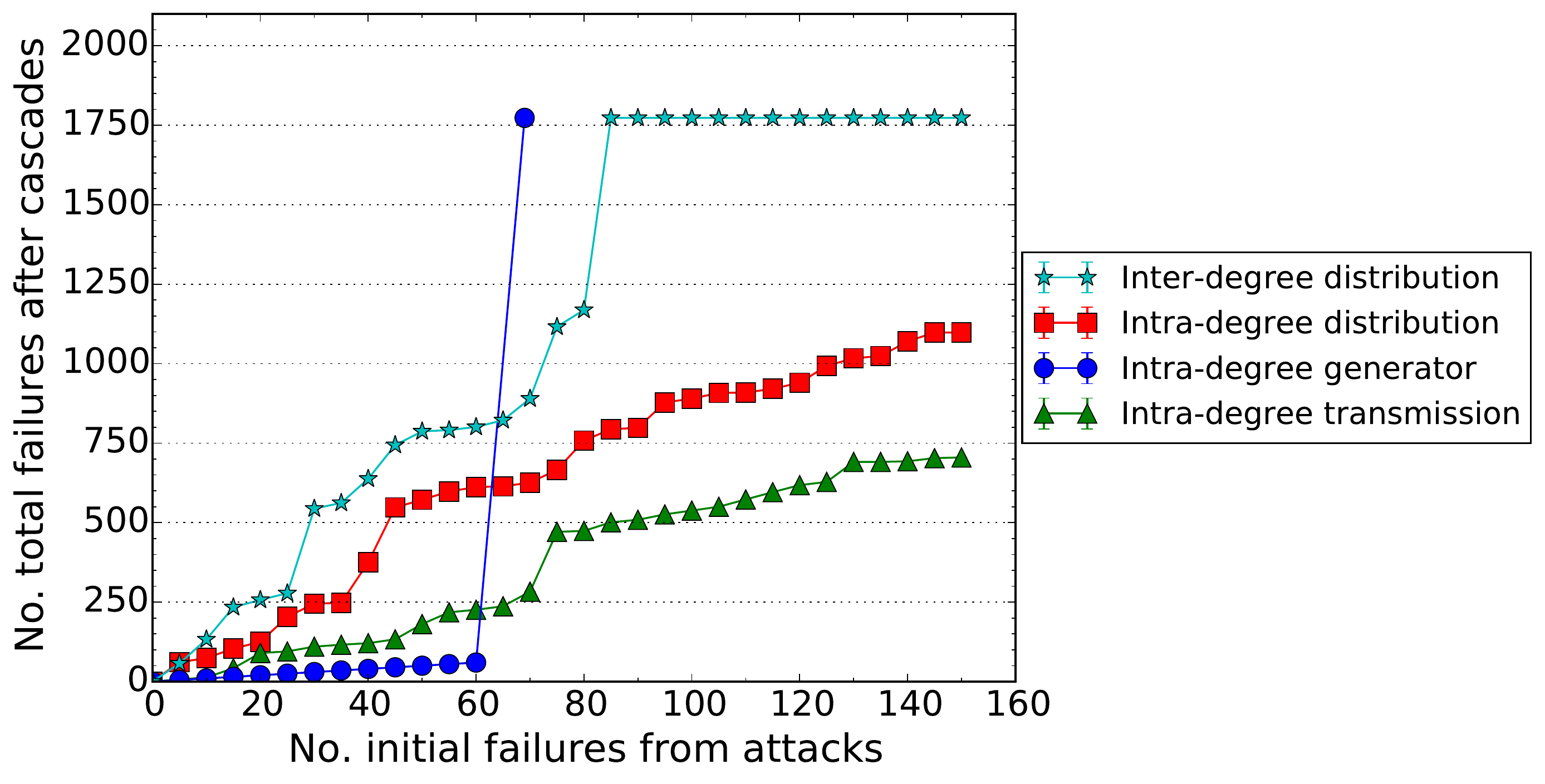}
	\hspace*{\fill}
	\caption{Targeted attacks on real networks.}
	\label{fig:exp_3_failure_cnt}
\end{figure}

Figure \ref{fig:exp_3_failure_cnt} shows the results of the experiments discussed above. As the figure points out, attacking distribution substations according to their inter-degrees is the most effective in terms of final failures. In fact, these attacks cause the failure of several relay nodes that depend upon those distribution substations, therefore limiting the access to the communications network and triggering further failures that overall generate devastating effects. The attack on distribution and transmission substations, based on intra-degrees, has less effects. Nevertheless, the attack on distribution substations has more impact due to their role in providing energy to the communications network. The attack on sole generators has no particular impact until there is at least one active generator. We recall that we have 69 generators in the real topologies that we consider. This shows a limitation of our approach that only considers connectivity to the generators as a sufficient condition for a power node to provide sufficient energy to the communications network. In reality, generators have a maximum production capacity that cannot be exceeded. We will consider these aspects in our future works.

\section{Related works}
The study of failure propagation in interdependent networks has received considerable attention in recent years \cite{gao_networks_2012, dong_percolation_2012, dong_robustness_2013, huang_robustness_2011, barthelemy_spatial_2010, rosato_modelling_2008, hu_percolation_2011, bashan_percolation_2011, son_percolation_2012, cellai_percolation_2013, parshani_critical_2011}. Buldyrev et al. introduced this new research area in their pioneering work \cite{buldyrev_catastrophic_2010}. In our work, we adopt the Uniform model described in this paper. In particular, an initial failure is generated by failing a fraction $1-p$ of nodes. Additional failures occur within networks, according to a criteria based on the giant component, and across network, due to the interdependencies.

The Uniform model has been extended in numerous ways, surveyed in \cite{gao_single_2014}. In particular, in \cite{shao_cascade_2011} nodes are allowed to have a random number of interdependencies. In \cite{parshani_critical_2011, cellai_percolation_2013, hu_percolation_2013} inter-edges are assigned according to an inter-similarity measure such that nodes with high intra-degrees tend to be dependent on each other. The authors of \cite{huang_robustness_2011, dong_percolation_2012} consider non-uniform targeted attacks, while the works in \cite{dong_robustness_2013-1, gao_robustness_2012, gao_networks_2012, son_percolation_2012} consider a generalization to the case of $n$ interdependent networks. Finally, a recent work \cite{huang_small_2015} proposes the small clusters (SC) model discussed in this paper. This model extends the analysis by considering clusters of nodes above a certain size as operational.

\section{Conclusion}
In this paper we proposed a realistic model for failure propagation in the interdependent power and communications networks. Our model, \OM{}, considers heterogeneity of network elements, complex logical and physical interdependencies, as well as complex failure propagation conditions. We experimentally compare \OM{} against two existing models. We adopted synthetic and real networks and considered random and targeted attacks. Results show that previous approaches overlooked several critical aspects of the system functionality, which often result in an underestimation of the failure propagation process. \OM{} provides the basis for future development of formal frameworks to predict the evolution of failures, as well as to analyze and improve their robustness.


\bibliographystyle{IEEEtran}
\bibliography{IEEEabrv,Towards_a_Realistic_Model_for_Failure_Propagation_in_Interdependent_Networks}

\end{document}